\documentclass[
    prx,
    reprint,
    longbibliography,
    nofootinbib,
    amsmath,amssymb,
    superscriptaddress,
    aps,
    floatfix,
]{revtex4-2}
\usepackage{xprintlen}
\usepackage[T1]{fontenc}
\usepackage{times}

\usepackage[table,usenames,dvipsnames]{xcolor}

\usepackage{hyperref}
\hypersetup{
  colorlinks=true,
  hypertexnames=false,
  linktocpage,
  colorlinks=true, 
  urlcolor=magenta!90!black,    
  linkcolor=blue!60!black,
  citecolor=black!60 
}

\usepackage{physics,siunitx}
  \usepackage{amsthm,amssymb,amsmath}
\usepackage{bbm,mathtools}

\usepackage{MnSymbol} 
\usepackage{youngtab}
\usepackage{quantikz}

\usepackage{booktabs}
\usepackage{enumitem}

\usepackage[capitalize,compress]{cleveref}

\usepackage{complexity}

\usepackage{pgfplots}
\usepackage{tikz}
\usepackage{pgffor}
\usepackage{algorithm}
\usepackage{algpseudocode}
\algrenewcommand\algorithmicrequire{\textbf{Input:}}
\algrenewcommand\algorithmicensure{\textbf{Output:}}

\usepgfplotslibrary{
  fillbetween,
  }

\usetikzlibrary{
  pgfplots.groupplots,
  arrows.meta,
  backgrounds,
  chains,
  decorations.pathreplacing,
  decorations.pathmorphing,
  fit,
  intersections,
  through,
  positioning,
  shapes,
  shapes.geometric,
  shapes.arrows,
  calc,
  3d
  }

\newclass{\stoqma}{StoqMA}
\newclass{\classP}{P}
\newclass{\bqp}{BQP}
\newclass{\qcam}{QCAM}
\newclass{\postbqp}{postBQP}
\newclass{\posta}{postA}
\newclass{\postiqp}{postIQP}
\newclass{\classa}{A}
\newclass{\bpp}{BPP}
\newclass{\fbpp}{FBPP}
\newclass{\pp}{PP}
\newclass{\cocp}{coC_=P}
\newclass{\ph}{PH}
\newclass{\np}{NP}
\newclass{\conp}{coNP}
\newclass{\gapp}{GapP}
\newclass{\approxclass}{Apx}
\newclass{\gapclass}{Gap}
\newclass{\sharpP}{\#P}
\newclass{\ma}{MA}
\newclass{\am}{AM}
\newclass{\qma}{QMA}

\newclass{\hog}{HOG}
\newclass{\quath}{QUATH}
\newclass{\bog}{BOG}
\newclass{\xeb}{XEB}
\newclass{\xhog}{XHOG}
\newclass{\xquath}{XQUATH}
\newclass{\maxcut}{MAXCUT}
\newclass{\sat}{SAT}
\newclass{\maxtwosat}{MAX2SAT}
\newclass{\twosat}{2SAT}
\newclass{\threesat}{3SAT}
\newclass{\sharpsat}{\#SAT}
\newclass{\se}{Sign Easing}
\newclass{\classx}{X}

\newcommand{\mc}{\mathcal}
\newcommand{\mb}{\mathbb}

\newcommand{\id}{\mathbbm{1}}
\newcommand{\ee}{\mathrm{e}}
\newcommand{\ii}{\mathrm{i}}

\newcommand{\bin}{\{0,1\}}

\newcommand{\QuICS}{
Joint Center for Quantum Information and Computer Science, NIST/University of Maryland, College Park, MD 20742, USA}
\newcommand{\simons}{
Simons Institute for the Theory of Computing, University of California at Berkeley, Berkeley, CA 94720, USA}
\newcommand{\ethz}{
Institute for Theoretical Physics, ETH Z\"urich, 8093 Zürich, Switzerland
}
\newcommand{\jqi}{
Joint Quantum Institute, University of Maryland, College Park, MD 20742, USA
}
\newcommand{\dqc}{
Duke Quantum Center, Duke University, Durham, NC 27701, USA
}

\definecolor{ingo}{rgb}{.8,.5,0}

\usepackage{xprintlen}

\begin{document}

\title{Blind calibration of a quantum computer}

\author{Liam M. Jeanette}
\email{liam.jeanette@duke.edu}
\affiliation{\jqi}
\affiliation{\dqc}

\author{Jadwiga Wilkens}
\affiliation{Institute for Integrated Circuits, Johannes Kepler University Linz, 4040 Linz, Austria}

\author{Ingo Roth}
\affiliation{Quantum Research Center, Technology Innovation Institute, Abu Dhabi, UAE}

\author{Anton Than}
\affiliation{\jqi}

\author{Alaina M. Green}
\affiliation{\jqi}

\author{Dominik Hangleiter}
\email{mail@dhangleiter.eu}
\affiliation{\simons}
\affiliation{\ethz}
\affiliation{\QuICS}

\author{Norbert M. Linke}
\affiliation{\dqc}
\affiliation{\jqi}

\date{\today}

\begin{abstract}
The calibration of quantum measurements is limited by the ability to accurately prepare  quantum states under unknown device errors.  
We develop an accurate calibration protocol for the measurement apparatus of a quantum computer that is `blind' to the state preparation.
Blind calibration quantifies and corrects measurement errors from simple tomographic data on a noisy quantum state. 
Importantly, it calibrates multiple error mechanisms in a single experiment, eliminating the need for bespoke, separate calibration experiments.
Using a trapped-ion quantum computer, we systematically demonstrate the accuracy of the method. 
We  use blind calibration to estimate the native calibration parameters of the experimental system. 
The recovered calibrations are consistent with directly measured values and perform similarly in predicting the state properties. 
\end{abstract}

\maketitle
\let\oldaddcontentsline\addcontentsline
\renewcommand{\addcontentsline}[3]{}

\section{Introduction}
	\label{sec:intro}

In order to build and calibrate large-scale, low-noise quantum computers, precise and accurate characterization techniques are needed. 
These techniques must be data-economical and light on post-processing.
Quantum state tomography \cite{hradil_quantum-state_1997,james_measurement_2001}, detector tomography \cite{luis_complete_1999,lundeen_tomography_2009} and process tomography \cite{chuang_prescription_1997,poyatos_complete_1997} are all designed to achieve this for different aspects of a quantum computing device. They are limited by the quality of their input data, which are generally only as precise and accurate as the measurement apparatus itself, a problem known as the \emph{calibration problem}~\cite{dariano_quantum_2004}.
At least a single high-quality component of an experiment, say high-quality measurement~\cite{dariano_quantum_2004} or state preparation~\cite{motka_efficient_2014,rehacek_operational_2010}, seems to be required in order to serve as a reliable probe to characterize errors and then improve performance based on that information. 

This is why it is desirable to solve the calibration problem in a different way. 
Consider the example of super-resolution spectroscopy techniques \cite{schmidt_multiple_1986,roy_estimation_1986}, where a fundamental limit---the Nyquist limit---is overcome by making an assumption about the signal, for instance, that it is composed of a single or few frequency components.
Then, only a very small number of measurements is required to obtain a highly precise estimate of the frequency of a signal. 

\begin{figure*}
    \includegraphics{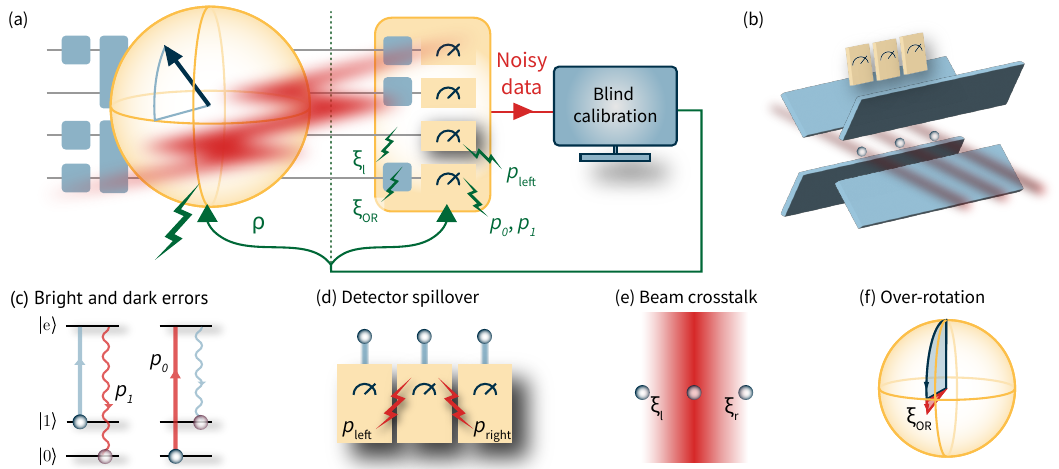}
    \caption{\textbf{Blind calibration of a trapped ion quantum computer.} (a) Since errors happen throughout a quantum computation, it is difficult to distinguish measurement errors from state preparation errors.
    The blind calibration protocol uses prior knowledge about the structure of the data in order to distinguish those errors in noisy data and thus allows us to accurately calibrate the device. 
    (b) Experimental setup. Ions are held in a linear Paul trap, probed using individual laser beams, and detected using individual detector channels. (c) Ion fluorescence on the transition between the computational state $\ket 1$ and an excited state $\ket e$  is used for readout. A bright error occurs when an ion in the bright $\ket 1$ state off-resonantly decays to the dark $\ket 0$ state with probability $p_1$ or vice versa a dark error with probability $p_0$. 
    (d) In a detector spillover error, a signal on one detector generates spurious photon counts on adjacent detectors due to electrical or optical crosstalk. This causes a false detection on the left detector with probability~$p_{\text{left}}$, and on the right detector with probability~$p_{\text{right}}$. 
    (e) A crosstalk error occurs, when the laser beam targeting a particular ion causes unintended rotation by an angle~$\xi_{r}(\xi_l) \cdot \frac \pi2$ on the neighboring right (left) ion. 
    (f) A target~$\frac{\pi}{2}$ operation over- or under-rotates by an angle~$\xi_{\text{\rm OR}}\cdot \frac \pi2$.}
    \label{fig:intro-fig}
\end{figure*}

Similarly, structural properties of the signal, which are known a priori, can be used to solve the calibration problem on a quantum computer and distinguish different types of errors in an experimental setup, see \cref{fig:intro-fig}(a). 
A prominent example of such a solution is gate set tomography~\cite{merkel_self-consistent_2013,blume-kohout_robust_2013,blume-kohout_demonstration_2017,brieger_compressive_2023}, where the measurements generated by a gate set are exploited in order to self-consistently perform accurate tomography of the gate set even in the presence of state preparation and measurement (SPAM) errors. 
Recently, a number of techniques have been proposed to characterize SPAM errors in order to subsequently calibrate tomographic measurements of a quantum computing device---a task that has been referred to as `self-calibrating'~\cite{
mogilevtsev_relative_2009,mogilevtsev_calibration_2010,mogilevtsev_self-calibration_2012,sim_proper_2019,branczyk_self-calibrating_2012}, `self-consistent'~\cite{moore_self-consistent_2020}, `simultaneous'~\cite{jayakumar_universal_2024}, or `blind'~\cite{roth_semi-device-dependent_2023} tomography. 
Other protocols solve the calibration problem in a way that mitigates measurement errors \cite{starek_measurement-device_2025,ma_corrupted_2024}, or model measurement errors in order to obtain more precise estimates of quantum instruments~\cite{stricker_characterizing_2022}.

Here, we experimentally demonstrate that the calibration problem can be solved, using a method we call \emph{blind calibration}. 
We focus on the problem of calibrating the \emph{measurement apparatus}, i.e., the quantum and classical operations used to perform measurements in arbitrary product bases. 
Using a small trapped-ion quantum computer, 
we show that a single set of simple projective Pauli measurement data can be used to first identify calibration errors and then infer their magnitudes. 
These values are then used in post-processing to correct for the miscalibrations. 

Importantly, our calibration technique is insensitive or \emph{blind} to the precise state 
used for calibrating the measurements.  
Moreover, many different calibration errors can be simultaneously calibrated in a single experiment.
In this way, blind calibration eliminates the need for multiple separate calibration experiments that each require high accuracy.

On the trapped-ion quantum computer, we use blind calibration to calibrate measurements suffering from single-qubit overrotation, beam crosstalk, classical readout, and detector spillover errors. 
Using any state from a set of suitable probe states, we identify the native system errors to the same level of accuracy as direct, custom-tailored measurements for each individual error source. 
The total number of experiments is comparable to that required for the direct measurements. 

Our work establishes that the calibration problem for the measurement apparatus can be solved not only in principle, but also in practice using a single `blind' experiment, with competitive performance compared to a combination of traditional techniques.

This paper is structured as follows. 
We begin by outlining the idea of blind calibration in \cref{sec:theory}, and the experimental setup and relevant miscalibrations in \cref{sec:experimental_setup_and_calibration_errors}. 
We then benchmark the method for those miscalibrations using numerical simulations and intentional miscalibrations in the experiment in \cref{sec:benchmarks}, demonstrating its efficacy. 
We then show our experimental results on calibrating the native system errors in \cref{sec:experimental_results}. We give a detailed comparison of accuracy and resource requirements compared to traditional, direct calibration techniques in \cref{sec:comparison_to_direct_calibration}. 
Finally, we conclude with a discussion and outlook in \cref{sec:conclusion}.

\section{Blind Calibration}
\label{sec:theory}

In blind calibration, we consider a tomographic set of measurement operators $\mc O$, following the approach of Ref.~\cite{roth_semi-device-dependent_2023}. 
Suppose, we measure an operator $M \in \mc O$. 
If there are miscalibrations with parameters $\zeta$, we effectively measure a different operator $ M(\zeta)$, 
resulting in data of the form 
\begin{align}
\label{eq:measurement data general}
y(\zeta,\rho) = \tr[ M(\zeta) \rho],
 \end{align} 
Our goal is to identify the calibration parameters $\zeta$ using measurements on suitable probe states $\rho$. 
The data \eqref{eq:measurement data general} can be decomposed as 
\begin{align}
\label{eq:linear measurement model}
    y(\zeta,\rho)= \sum_{j} \xi_j(\zeta) \tr[N_{j} \rho], 
\end{align}
in terms of operators $N_{j}$ weighted by calibration coefficients $\xi_j(\zeta)$, see Appendix~\ref{app:algorithm} for details. For small $\zeta_i \ll 1$, we can always reduce the number of calibration coefficients $\xi_j$ to at most $k +1$ linearly independent coefficients, where $k$ is the number of calibration parameters~$\zeta_i$. 

There are two cases of particular interest. 
The first case is a coherent error, when $M(\zeta) = \ee^{-\ii \zeta G/2} M \ee^{\ii \zeta G/2} $ for some rotation generator $G$. 
If $M,G$ are anti-commuting Pauli operators we find  $M(\zeta)= \cos(\zeta) M + \ii \sin (\zeta) MG$, which we can linearize as $M(\zeta) \approx M + \ii \zeta MG$ for small errors. 
In this case we therefore have $\xi_0 =1$ and $\xi_1 = \ii \zeta$. 
The second case is an incoherent readout error, which we model as a stochastic matrix~$S$ acting on the computational-basis measurement outcomes $b \in \bin^n$, $\proj b \mapsto \sum_{c \in \bin^n} S_{bc} \proj c$. If a measurement is performed in a basis different from the computational basis, the states $\ket c$ should be understood as the $\pm 1$- eigenstates $\ket b_M$ of the corresponding measurement operator~$M$. 

To infer $\zeta$ from the measurement data $y$, we find the best fit of the signals $\xi$ and $\rho$ given the measurement data using alternating minimization over the valid signals, characterized by rank-1 quantum states for $\rho$ and valid assignments of $\xi$. 
Given~$\xi$, we can then solve a linear system to obtain the calibration parameters $\zeta$. 
Given $\zeta$, we can then calibrate future measurements by including the miscalibration in the corresponding data model, and correcting for it in post-processing. 
Here, we do so using the same data for the task of state tomography, which is the most general task since the full description of the quantum state characterizes all possible measurements. 

The accuracy in the recovery of  a given set of calibration parameters depends on the choice of state.
This can be explained through \emph{slack} in the dependence of the objective function---i.e., the deviation of the measurement data from the measurement model \eqref{eq:linear measurement model}---on the calibration parameters for different states.
Take for example, a $\ket 0$ state and a rotation around the $Z$-axis. The resulting data will be independent of the value of the rotation, and its overrotation in particular, and therefore we cannot expect to obtain a good recovery of the corresponding calibrations. 
In general, the sensitivity landscape will be significantly more complicated. Therefore, to obtain an optimal recovery in terms of precision and accuracy, the probe states should be selected carefully for a given error model. 
We expect nonetheless that random states will yield accurate recoveries with high probability since no slack directions in state space are distinguished. 
One can thus consider a two-step calibration procedure in which in the first step, the relevant miscalibrations are identified using random states, and in a second step, they are calibrated to high accuracy using optimal probe states. 


Our blind calibration protocol is closely related to the blind tomography task, introduced in Ref.~\cite{roth_semi-device-dependent_2023}. There, the goal is to simultaneously infer $\rho$ and $\zeta$ from the data vector $y$, and it was proven that  this problem can be solved in principle for an idealized calibration model, assuming that the state preparation is approximately pure. 
The algorithms and theory developed in Ref.~\cite{roth_semi-device-dependent_2023} focus on state reconstruction in an idealized setting. 
In contrast, we here want to obtain a precise estimate of the device calibration $\zeta$ using an imperfectly prepared but suitable probe state $\rho$ in a real system.
To achieve this, we adapt and improve the protocol and algorithm. 
We arrive at a practically applicable \emph{blind calibration} method that incorporates all a priori know structural properties of the signal and makes use of projective measurements.

\begin{figure*}
    \centering
    \includegraphics{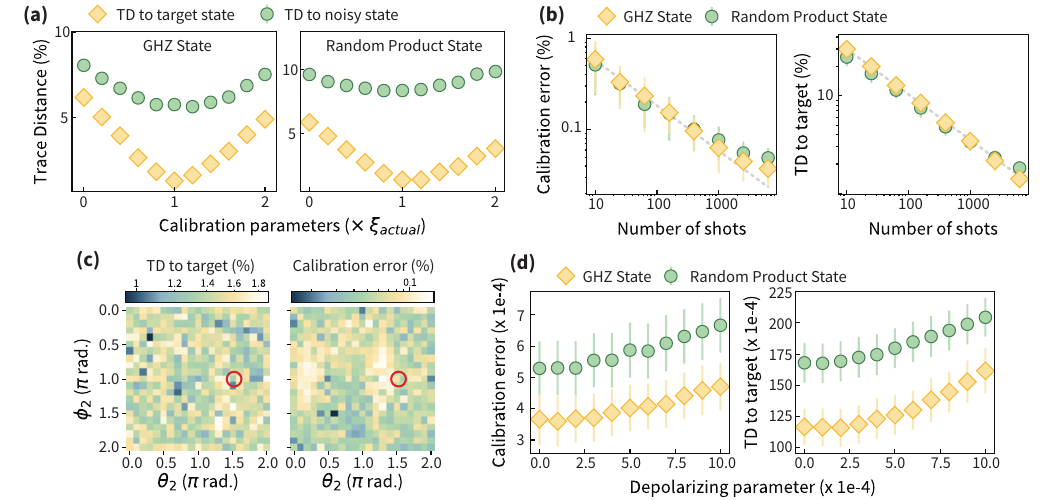}
    \caption{\textbf{Numerical benchmarks of blind calibration} using a three-qubit GHZ state and a randomly-drawn three-qubit product state. 
    We simulate the four error types discussed in the main text, resulting in six calibration parameters. 
    We set the parameter values to $\xi_{\text{actual}}$ given in \cref{eq:xi actual} of Appendix~\ref{app:error-models}. 
    (a) Evidence for validity of trace distance to the target state as a measure of success.
    Comparison of the state estimate obtained from calibrated tomography with calibration parameters $(1-c) \xi_{\text{ideal}} + c \xi_{\text{actual}}$ for $c \in [0,2]$ with $\xi_{\text{ideal}} = (1,0, \ldots)$ and $\xi_{\text{actual}}$ given in \cref{eq:xi actual} using $10000$ simulated shots per measurement basis from noisy state preparations using a Pauli noise channel with strength 1\% per gate for the GHZ state (left) and the product state (right).  
    Shown are the TD to the noisy (simulated) state preparation (diamonds) and the target state (circles). Error bars are one standard deviation.
    (b) Calibration error (left) for the seven recovered parameters. The recovered parameters are used to perform calibrated state tomography, with the resulting trace distances to the target (right).
    Both calibration error and trace distance are shown as a function of the number of experimental shots per basis with one standard deviation error bars, decaying as an inverse square root (dashed line). 
    (c) State sensitivity for the product state. The two outer qubits are held at fixed angles, while the middle qubit is scanned through all possible states. The product state is shown by the red circle. The trace distance to the target and the calibration error remain constant to within a percent.
    (d) Calibration parameter recovery with increasing local depolarizing noise (left). 
    Local depolarizing strength is chosen based on available noise models of similar hardware \cite{ionq_noise_model}. 
    The recovered parameters are used to perform calibrated state tomography (right). 
    Calibration error is insensitive to depolarizing noise to within a tenth of a percent, which translates to insensitivity in trace distance to within half a percent.
    }
    \label{fig:simulated-benchmarking}
\end{figure*}

\section{Experimental setup and calibration errors}
\label{sec:experimental_setup_and_calibration_errors}

We use a trapped-ion quantum computer consisting of a linear chain of $^{171}$Yb$^+$ ions trapped in a linear Paul trap as shown in \cref{fig:intro-fig}(b).  Qubit states $\ket{0}$ and $\ket{1}$ are encoded in the hyperfine states of the ground level, $\ket{^2S_{1/2}, F = 0, m_F = 0}$ and $\ket{^2S_{1/2}, F=1, m_F=0}$, respectively. Each qubit is initialized to $\ket{0}$ using optical pumping  \cite{olmschenk_manipulation_2007}, and coherent operations are applied to prepare an arbitrary state using individually addressed Raman beams \cite{debnath_demonstration_2016}. The state of each ion can be read out individually in the qubit $Z$-basis using state-dependent fluorescence, in which ions in the $\ket{1}$ state emit photons that are counted with a multi-channel photomultiplier tube (PMT), and ions in $\ket{0}$ remain dark.

For blind calibration, we perform projective measurements in all Pauli bases. In order to implement those measurements, we apply $\pi/2$-single-qubit rotations around the $X$ or $Y$ axis, $R_X(-\pi/2)$ or $R_Y(\pi/2)$, respectively, before the native measurement in the computational $Z$-basis. 
%
We then model the dominant sources of measurement errors in the trapped-ion system and calculate the corresponding transformations of each Pauli basis required for the model \eqref{eq:linear measurement model}. 
We calibrate the four dominant measurement errors in our setup: overrotations, beam crosstalk, dark and bright readout error, and detector spillover. 

When applying single-qubit basis rotations, $R_X(\pi/2)$ and $R_Y(-\pi/2)$,
there is a possibility of \emph{over- or underrotating} the qubit by an angle of $ \pm \xi_\text{\rm OR} \cdot \pi/2$, shown in \cref{fig:intro-fig}(f). 
Assuming the angle of this excess rotation is small, it results in an unwanted projection into the $Z$ basis, leading to transformed Pauli observables 
\begin{align}
    X &\mapsto X - \xi_\text{\rm OR} \pi Z \\
    Y &\mapsto Y - \xi_\text{\rm OR} \pi Z.
\end{align}
Since no single-qubit rotations are performed to measure in the native $Z$ basis, only the $X$ and $Y$ bases are affected by overrotation errors.

To perform state readout, we excite a cycling transition from $^2S_{1/2}(F=1) \rightarrow{} ^2P_{1/2} (F=0)$. 
Ions in $\ket{1}$ fluoresce, while ions in $\ket{0}$ remain dark.
During this process, the laser can off-resonantly drive a transition that causes decay into the opposite state with probability, causing a readout error.
We call this \textit{dark error} or \textit{bright error}, occurring with probabilities $p_0$ and $p_1$, respectively, as shown in \cref{fig:intro-fig}(c). 
Under a dark error, the $\ket 0$-state projector transforms as 
\begin{align}
    \proj{0}_M &\mapsto (1-p_0)\proj{0}_M + p_0\proj{1}_M
\end{align}
in all Pauli bases $\ket b_M$.

In Appendix~\ref{app:error-models}, we detail the full measurement model including also \textit{detector spillover}, where a bright ion spills to neighboring detector channels, and \textit{beam crosstalk}, where light from the laser beam causes unwanted illumination on neighboring ions.

\begin{figure*}
    \centering
    \includegraphics{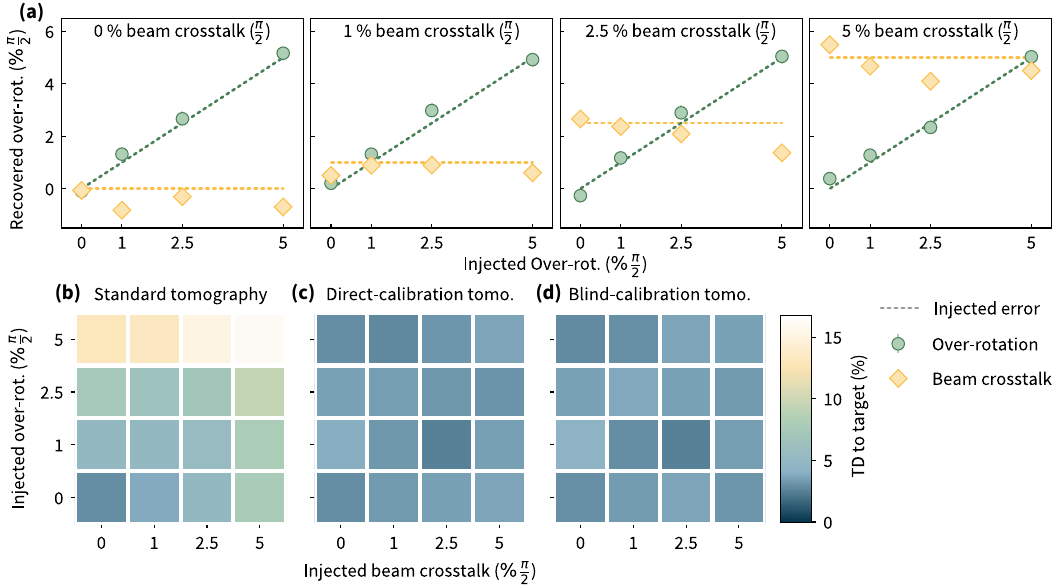}
    \caption{
    \textbf{Recovery of intentional miscalibrations.}
    In order to demonstrate the experimental viability of blind calibration, we intentionally inject overrotation and nearest-neighbour beam crosstalk errors (with $\xi_r = \xi_l$) of varying magnitude $[0,1,2.5,5]\cdot  \%\frac \pi2 $ by performing appropriate single-qubit rotations, when taking tomographic data of a randomly drawn 3-qubit product state in the $X$-$Z$ plane with polar angles given by $\theta_1=1.237\pi,\theta_2=0.670\pi,\theta_3=1.823\pi$. 
    (a) Blind-calibration estimates of the overrotation (green circles) and crosstalk errors  (yellow diamonds) compared to the injected errors (green/yellow dashed lines) as a function of the injected overrotation for different values of injected beam crosstalk. 
    (b) Assessment of the quality of  different calibrations using the trace distance (TD) to the target state of the state estimate obtained from standard tomography, direct-calibration tomography, and blind-calibration tomography. }
    \label{fig:injected-error-fig}
\end{figure*}


\section{Benchmarking blind calibration}
\label{sec:benchmarks}

To benchmark the blind calibration protocol, we numerically simulate quantum circuits and perform experiments with intentional miscalibrations. 
We focus on three-qubit states, the minimum number of qubits to see the multi-qubit errors described above. 
Specifically, we choose two paradigmatic states: a randomly-drawn product state and a GHZ-state. 

We 
quantify the quality of the recovered calibrations in the \emph{calibration error} 
\begin{align}
    E(\xi, \tau) \coloneqq  \frac 1 k \sum_{j=1}^k | \xi_j - \tau_j| , 
\end{align}
measured by the normalized $\ell_1$-norm between the recovered calibration parameters $\xi$ and the true parameters $\tau$. 
In order to assess the quality of the calibration parameters inferred from blind calibration when the true parameters are unknown, we perform state tomography on the same type of experimental data using the corrected measurements, see Appendix~\ref{app:tomographies} for details.
We then compute the trace distance (TD) of the state estimate~$\rho$ to the target state~$\psi$, 
\begin{align}
d_{\tr}(\rho, \psi) = \frac 12 \tr(|\rho - \psi|), 
\end{align}
  as a figure of merit for the quality of the calibration, in place of the \emph{unknown} distance to the experimentally prepared state.

\subsection{Numerical simulations}

In a first step, we benchmark the protocol by simulating miscalibrated projective measurements. 

In \cref{fig:simulated-benchmarking}(a), we first demonstrate  that the TD to the target state is indeed a faithful comparative indicator of the quality of the recovered calibration parameters.
\cref{fig:simulated-benchmarking}(b) shows the performance of blind calibration when recovering six calibration parameters (beam crosstalk, left detector spillover, right detector spillover, dark error, bright error, and overrotation parameters) using simulated data in terms of the calibration error and TD to the target state. 
Because there are no state preparation errors, any deviation from the target state is due to shot noise and calibration errors. 
As we increase the number of simulated shots, decreasing the statistical uncertainty, we see that the deviation from the target state and expected errors both tend toward the expected result following an inverse square root decay. 
This shows that the miscalibrations are recovered to sufficient accuracy so that they do not limit the state recovery.
We are able to recover calibration errors to well within 1\% using only 1000 shots per measurement basis. 
For the six modeled errors, we assume that randomly drawn product states will provide adequate accuracy while remaining robust to state preparation noise. We show in \cref{fig:simulated-benchmarking}(c) the randomly-drawn three-qubit product state's surrounding landscape. We scan the state of the middle qubit while keeping the outer qubits fixed. For each state, we simulate $10^4$ measurement shots and recover a set of calibration parameters for which the calibration error is shown. We use these calibration parameters to perform calibrated-tomography for each state, measuring the trace distance to the target. For this combination of state and errors, there is minimal change in both calibration error and trace distance as the state is varied. Therefore, we conclude that this state is adequate for calibration under these circumstances.

Finally, we verify the the blindness of our procedure to state-preparation errors by computing the sensitivity of the calibration recovery from this randomly-drawn product state through simulations of noisy state preparations. We prepare the product state with varying degrees of depolarizing noise, then recover the set of calibration parameters which we use to perform calibrated tomography. 
The calibration error and trace distance as a function of depolarizing strength are shown in \cref{fig:simulated-benchmarking}(d). We see that the recoveries of the calibration parameters only change by roughly $10^{-4}$ as the depolarizing strength is varied by $10^{-3}$. Thus, they are only very weakly sensitive to the induced noise.
In comparison, the direct calibration recovery---for the example of SPAM recovery---changes by half the depolarizing error, constituting half of an order of magnitude improvement of blind versus direct calibration.  
At the same time, the trace distance to the target state remains a good indicator of the calibration accuracy.

\begin{figure*}
    \centering
    \includegraphics{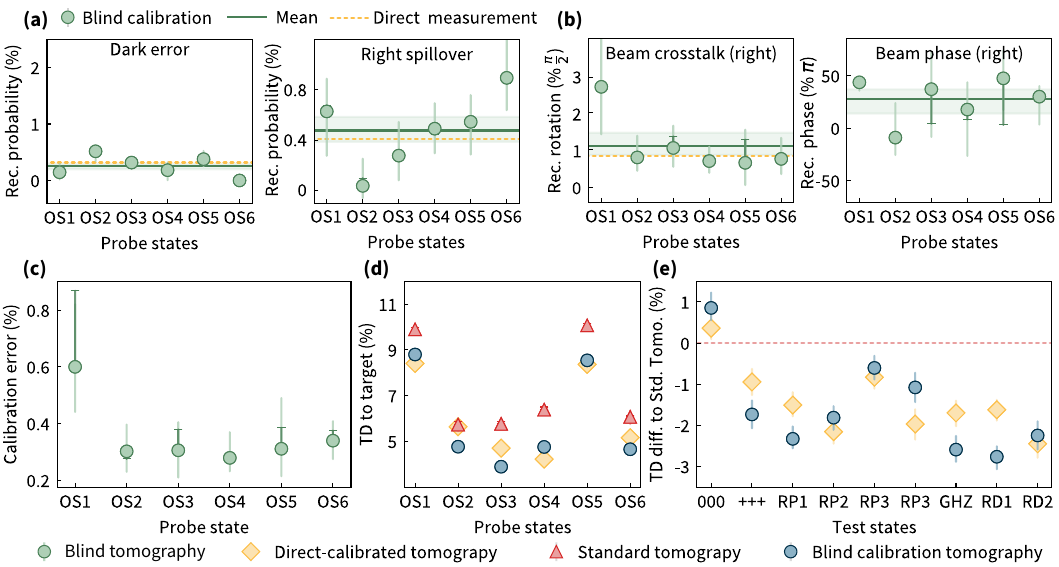}
    \caption{\textbf{Blind calibration of the native errors of the trapped-ion quantum computer} using six optimal \emph{probe states} labeled OS1 to OS6, see Appendix~\ref{app:optimal states} for details. Blind-calibration estimates of the calibration parameters are represented by green circles. 
    Error bars are composed of the systematic error (dark green with caps) and one standard deviation statistical error (light green), see Appendix~\ref{app:error-bars} for details. The average of the estimated calibration parameters over the six probe states and its standard deviation are represented by a solid green line and shaded area, respectively, and the estimates from the direct measurement by dashed yellow lines. 
    (a) Estimates of dark error $p_0$ and right detector spillover error $p_{\text{right}}$. 
    (b) Estimates of right beam crosstalk magnitude $\xi_{r}$ and right beam crosstalk phase $\phi_{r}$.
    (c) Calibration error $E(\xi, \delta)$ of the blind calibration ($\xi$) compared to direct measurements ($\delta$). 
    (d,e) To assess the quality of different calibrations, we compute the distance (TD) to the target state of the state estimates obtained from standard tomography (triangles), direct-calibration tomography (diamonds), and blind-calibration tomography (circles). 
    For blind-calibration tomography, we use the averages obtained from the six optimal probe states as our calibration parameter estimates.
    In (d), we show the quality of the estimates for tomographic data from the six probe states OS1--OS6. 
    (e) To demonstrate that the blind calibration generalizes beyond the probe states, we also obtain tomographic data from several independent \emph{test states}: the basis state $\ket{000}$, the Hadamard basis state $ \ket{+++}$, four random product states labeled RP1 to RP4, the GHZ state, and two states generated by random deep circuits, labeled RD1 and RD2, see Appendix~\ref{app:optimal states} for details. 
    We show the difference $d_{\tr}(\rho^{*}, \psi) - d_{\tr}(\rho, \psi)$ of the trace distances between the standard tomography estimate $\rho$  and  the direct/blind calibration estimate $\rho^{*}$ to the target state $\psi$.
    In this representation, standard tomography achieves a value of $0$ (solid line), and the more negative the values of direct/blind-calibration tomography are, the closer the corresponding state estimate is to the target state.
    }
    \label{fig:native-error-fig}
\end{figure*}

\begin{figure}
    \centering
    \includegraphics{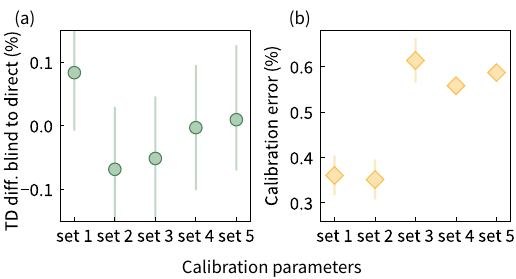}
    \caption{\textbf{Calibration parameter optimization.} We find an optimal set of parameters to include in the blind calibration model. Each set contains decreasing numbers of parameters. Set 1 includes the full set of 9 calibration parameters: overrotations, beam crosstalk magnitude, beam crosstalk phase, detector spillover, and dark and bright errors. Set 2 removes overrotations. Set 3 removes beam crosstalk phase from Set 2. 
    Set 4 removes beam crosstalk magnitude from Set 3. Set 5 removes detector spillover from Set 4, and thus only dark and bright errors remain. We perform rank-1 blind calibration for each set and show the calibration error in (b). We then perform rank-8 blind tomography of the test states and compare to direct tomography. The differences in trace distances are shown in (a). We conclude that removing overrotation from the model leads to an optimal recovery.
    }
    \label{fig:parameter-removal}
\end{figure}

\subsection{Intentional experimental miscalibrations}

We then verify the correct functioning of blind calibration in an experimental system by intentionally miscalibrating our system and recovering the \emph{known} miscalibrations. 
To this end, we prepare a three-qubit random product state and add varying amounts of overrotation and symmetric beam crosstalk with uniform phase to the system, resulting in two total error parameters. For overrotation errors, we add a small extra angle to the target qubit, while for beam crosstalk errors, we add small rotations to the neighboring qubits. 
We add rotations of 0\%, 1\%, 2.5\%, and 5\% (of $\frac{\pi}{2}$) and correct the data for classical readout errors (see Appendix~\ref{app:spam-measurement}) to ensure only rotational measurement errors are present. 
We perform $10^4$ experimental shots then use blind calibration to recover the injected errors, shown in \cref{fig:injected-error-fig}(a). 
We find accurate recoveries of the injected miscalibrations.

To assess the quality of the blind calibration estimates, we perform \emph{blind-calibration tomography}, where we use the parameters obtained in blind calibration, \emph{standard tomography}, where the calibration is assumed to be perfect (i.e. $\xi = 0$), and \emph{direct-calibration tomography}, where the calibration parameters are those obtained from direct measurements on the device.

In order to perform direct-calibration tomography, We directly measure readout errors and beam crosstalk magnitude. Readout errors can be estimated by measuring a single target ion in the $\ket{1}$ ($\ket{0}$) state. Any dark (bright) measurement on the target detector channel indicates a dark (bright) error. Additionally, any bright measurement on the neighboring channels indicates a spillover error. Beam crosstalk is measured by performing a Rabi flop on a target ion and observing its neighboring ions. It is important to note that in this system, due to drifts in the beam position, the Rabi frequency of each ion may shift, giving rise to innacurate single-qubit pulses. We routinely calibrate single-qubit gates by applying a $\pi/2$-pulse to each ion and correcting the duration of the pulse based on the result. In our method, we model slow-drifting systematic overrotations which may arise due to over or under-compensation during this calibration.

The results for the intentionally miscalibrated setup are shown in \cref{fig:injected-error-fig}(b-d), respectively.  
They demonstrate that, as expected, the quality of standard tomography (b) decreases with the amount of miscalibration.
In contrast, when using the directly measured (i.e., in this case, the injected) calibration parameters (c) and the parameters obtained from blind calibration (d) to inform tomography, we find that the quality of the state recovery remains constant independent of the amount of miscalibration.
Blind calibration also matches the performance of the direct calibrations in terms of the TD to the target state. 

Altogether, our numerical and experimental benchmark establish that blind calibration can faithfully calibrate multiple different error sources  in the presence of state-preparation errors from a single set of measurement data.  


\section{Native device calibration}
\label{sec:experimental_results}

We use blind calibration to calibrate the native errors of the system. 
These are  smaller than the injected errors, namely, on the order of $\lesssim 1 \%$, see \cref{fig:native-error-fig}. 
Hence, they require significantly higher-precision estimates.
In fact, the random product states chosen for the numerical benchmarks do not yield  sufficiently accurate and precise estimates of the calibration parameters, and we optimize probe states for the expected set of errors using simulated data, see Appendix~\ref{app:optimal states} for details. 
We find 6 different optimal states (with respect to different measures of optimality) and use them as probe states. 
We then measure a set of tomographic data using $10^4$ experimental shots per measurement basis. 

Given the data, to maximize accuracy, we compare how many and which calibration errors should be included in our model to obtain the best fit with the data, see \cref{fig:parameter-removal} and Appendix~\ref{app:full recovery} for details. 
We find that dark and bright errors, detector spillover and beam crosstalk should be included. 
The resulting calibration parameters for two readout errors and two crosstalk errors are shown in \cref{fig:native-error-fig} (a) and (b), respectively. 
The full set of estimated calibration parameters is shown in Appendix~\ref{app:full recovery}. 
We then take the average of the estimated calibration parameters over the six optimal probe states as our calibration.
The resulting parameters show close agreement with the best estimates from direct experimental diagnostics, where available.
We explain these traditional methods in Appendix~\ref{app:direct measurement}.

We assess the quality of blind calibration in comparison to  direct calibration. 
To this end, we again turn to the TD to the target state of the tomographic estimates of the states used in calibration, see \cref{fig:native-error-fig}(d). 
Due to the increased number of parameters in blind calibration, there is a risk for overfitting as compared to standard calibration. 
Therefore, we test how well the calibration generalizes to different settings, in particular, other state preparations.
We do so by comparing the quality of the calibration obtained from blind calibration on the optimal probe states with  those obtained from direct experimental diagnostics for the recovery of various new state preparations in \cref{fig:native-error-fig}(e). 
We find that the calibration parameters from blind calibration yield state estimates slightly closer to the target than uncalibrated standard tomography.  
This shows that the parameters obtained from blind calibration generalize to independent experiments on different states and thus are not an artifact of overfitting.
They also perform similarly well compared to the calibration parameters obtained from direct diagnostics, demonstrating the efficacy of blind calibration. 

\section{Comparison to direct calibration}
\label{sec:comparison_to_direct_calibration}
In our experimental implementation, we have shown that blind calibration can be used to calibrate a trapped-ion system with precision comparable to direct calibration. In other words, we have developed a single calibration method which performs equally well as using multiple tailored methods. Further, we achieve comparable calibration performance to direct calibration without requiring additional overhead. We measure $10^4$ experimental shots for each Pauli basis, or $2.7\times10^5$ total measurement shots. In comparison, direct calibration consists of SPAM correction and beam crosstalk calibration. SPAM correction requires $8\times10^4$ measurement shots per ion, while beam crosstalk requires $5\times10^4$ measurement shots per ion. In total, direct calibration required $6.3\times10^5$ total measurement shots for our experiment as compared to $2.7\times10^5$ shots for blind calibration, see \cref{tab:comparison} for the comparison.

\begin{table}
    \centering
  \begin{tabular}{l >{\ }c<{\quad}  c  >{\quad}c }
     & Tom.\ improvement & Shots & Scaling 1D chain  \\ 
      &   (\%) & ($\times 1000$) &  ($\times 1000$)  \\
     \hline\hline
     Blind  & 1.6(1) & 270 &  270 $(N-2)$ \\
     Direct & 1.5(1) & 630 &  210 $N$
    \end{tabular}
    \caption{
    \textbf{Comparison of blind calibration to direct calibration.} 
    We report the total number of experimental shots necessary to perform blind calibration and direct calibration for a single three-qubit state. We show the improvement in tomography accuracy, reporting the difference in average TD to the target between each method and standard tomography. A greater tomography improvement indicates a more accurate calibration as described in Appendix~\ref{app:tomographies}. Finally, we report the scaling of each calibration method with the number of qubits,~$N$. }
    \label{tab:comparison}
\end{table}

We propose a method for scaling blind calibration beyond three qubits. Due to typical Gaussian or $1/r$ scalings, it is assumed that nearest neighbor interactions will dominate the multi-qubit errors present in a given system. A nearest-neighbor interaction may act asymmetrically, requiring special treatment of ``left'' and ``right'' errors for a chain of ions. Therefore, to determine nearest-neighbor multi-qubit errors, we only need at most three qubits. However, for a long ion chain, errors may be non-uniform across the chain. Instead of measuring a complete tomographic data set for the entire chain, we can break the chain into sub-groups of three qubits, measuring the multi-qubit errors for each subgroup to construct a full calibration of the chain. This reduces the required measurements from exponential to linear, allowing for a scalable calibration protocol. This method may be extended to next-nearest-neighbor interactions, retaining a linear scaling.

We show in \cref{fig:simulated-benchmarking}(c,d) that the performance of blind calibration does not suffer due to noisy state preparations, allowing us to calibrate the system without having to rely on highly-accurate states. This is not generally true for direct calibration methods. For example, when characterizing readout errors, the accuracy of the estimate relies heavily on precise state preparation. In fact, we show in Appendix~\ref{app:direct-sensitivity} that in the presence of depolarizing noise, the calibration error for dark and bright error estimates scales linearly with the depolarizing strength.   

\section{Discussion and outlook}
	\label{sec:conclusion}

In this work, we develop a method to calibrate the measurement apparatus of  a quantum computer that is blind to a specific input state and its accuracy, and thus provides a solution to the calibration problem~\cite{dariano_quantum_2004}. 
Blind calibration requires only a single tomography experiment and thus offers an easy-to-use and data-efficient method for calibrating many error mechanisms. 
Thus, it eliminates the need for several tailored experiments that address each error one-by-one. 
At the same time, it achieves similar precision and resource requirements compared to traditional direct calibration techniques. 

An important advantage of blind calibration is that 
the calibration can be done \emph{post hoc} in a ``measure-first-think-later'' 
mindset similar to classical shadows \cite{huang_predicting_2020,helsen_estimating_2022-1,elben_randomized_2023}.
If an experimenter is unsure about the dominant sources of error affecting their system, they can simply take a set of tomographic data and then compare how well different error models describe the data in post-processing. 
In fact, when we performed blind  calibration of the native device errors (\cref{fig:native-error-fig}), only after analyzing the experimental data did we realize that different calibration parameters are required to account for beam crosstalk errors and that overrotation errors should be excluded from the model (see Appendix~\ref{app:parameter-optimization}). 
A systematic way to do this was sketched in Ref.~\cite{roth_semi-device-dependent_2023}, using sparse recovery techniques. 

Our work raises several  questions that we hope will be addressed in the future. 
First,  
our calibration experiments use data from three qubits, which is enough to calibrate measurement errors with at most next-nearest-neighbor interactions, assuming consistent errors across qubits. 
Probe states supported on more qubits, can be potentially used to detect a larger number of dominant error mechanisms. 
Second, 
when modeling the measurement errors in the ion-trap system, 
we expanded them to linear order in the calibration parameters $\xi$. 
In our setting, this approximation was sufficient. 
Our solution to the blind calibration problem using alternating minimization can be straight-forwardly adapted to nonlinear dependencies on the calibration parameter further extending the applicability of the method. 
Third, 
blind calibration is a general framework and can be applied to any physical platform and its specific error profile. 
Future work should establish the limits to the number and types of calibration errors that can be simultaneously recovered. 
Finally, we limited our optimization over different probe states to product states 
in order to restrict the optimization space. 
An intriguing problem that we leave open is to develop a detailed theory of sensitivity for blind calibration. 
Identifying properties of good probe states for certain error models would allow us to cheaply identify suitable states sensitive to those errors as the probe system is scaled up.

\begin{acknowledgments}

This material is based upon work supported by the U.S. Department of Energy, Office of Science, National Quantum Information Science Research Centers, Quantum Systems Accelerator (DE-FOA-0002253). Additional support is acknowledged from the National Science Foundation, Quantum Leap Challenge Institute for Robust Quantum Simulation (OMA-2120757) and Software-Tailored Architecture for Quantum Co-Design (STAQ) Award (PHY-2325080). 
DH is grateful for support from the Simons Institute of the Theory of Computing, supported by DOE QSA, and from the Swiss National Science Foundation through Ambizione Grant No.\ 223764.
JW has been supported by the project High-Performance Integrated Quantum Computing (FFG 897481) within Quantum Austria.
The authors thank Yingyue Zhu for support with the experiments, and Michael Straus and Nhung Nguyen for helpful discussions.
\end{acknowledgments}


\bibliographystyle{./myapsrev4-1}
\bibliography{doms_refs,more_refs}


\appendix
\clearpage
\let\addcontentsline\oldaddcontentsline%
\renewcommand{\tocname}{Appendices}
\tableofcontents

\section{Blind calibration algorithm}
\label{app:algorithm}

\subsection{Theoretical framework}

Consider a tomographically complete set of measurement operators $\mc O$. In the miscalibrated scenario, instead of $M \in \mc O$ we thus effectively measure an operator $ M(\zeta) = \sum_{M' \in \mc O} \xi_{M \rightarrow M'}(\zeta) M'$. 
Here, the $\xi_{M \rightarrow M'}(\zeta)$ are 
linear coefficients obtained when decomposing $M(\zeta)$ into the tomographically complete set $\mc O$. 
We can now group terms with the same $\xi_{M_i \rightarrow M'}(\zeta)\eqqcolon \xi_j(\zeta)$ into operators $N_{i,j}$, and rephrase the blind tomography model as a linear sum
\begin{align}
\label{eq:linear measurement model appendix}
    y_i(\zeta,\rho)= \sum_{j} \xi_j(\zeta) \tr[N_{i,j} \rho] \equiv \mc A(\xi, \rho)_i.
\end{align}
 Linearizing the dependence on $\zeta$ we can always reduce the number of calibration coefficients $\xi_j$ to at most $k +1$ linearly independent coefficients. 
We set $N_{i,0} = M_i$ to be the ideal target measurement. Thus, when the device is already perfectly calibrated, the linear coefficients are $\xi_0 = 1$ and $\xi_j = 0$ for all $j > 0$.

There are two cases of particular interest. 
The first case is a coherent error, when $M(\zeta) = \ee^{-\ii \zeta G/2} M \ee^{\ii \zeta G/2} $ for some rotation generator $G$. 
If $M,G$ are anti-commuting Pauli operators we find  $M(\zeta)= \cos(\zeta) M + \ii \sin (\zeta) MG$ 
, so that $\xi_{M \rightarrow M} = \cos(\zeta)$ and $\xi_{M \rightarrow \ii MG} = \sin(\zeta)$, which we can linearize as $M(\zeta) \approx M + \ii \zeta MG$ for small errors. 
The second case is an incoherent readout error, which we model as a stochastic matrix~$S$ acting on the computational-basis measurement outcomes $b \in \bin^n$, $\proj b \mapsto \sum_{c \in \bin^n} S_{bc} \proj c$. If a measurement is performed in a basis different from the computational basis, the states $\ket c$ should be understood as the $\pm 1$- eigenstates $\ket b_M$ of the corresponding measurement operator $M$. 

In order to further improve the precision of blind calibration, we use projective measurements $M_i = M_i^2$ rather than Pauli expectation-value estimates as done in  Ref.~\cite{roth_semi-device-dependent_2023}. 
Projective measurements in all $3^n$ Pauli bases (with $2^n$ possible outcomes each) describe the natively available measurement data and contain more information than expectation-value measurements of Pauli observables in the finite-sample regime. 
The two models can be translated into one another using the relation $\proj b_{M} = (\id + (-1)^b M)/2$, where $M$ is a Pauli matrix and $b \in \bin$. 
Below, we demonstrate the improvement in the precision of the recoveries.


To solve the blind calibration problem with high precision in practice, we minimize the bilinear objective function
\begin{align}\label{eq:tomography least squares}
    f(\xi, \rho) &= \frac{1}{2} \|\hat y-\mathcal{A}(\xi, \rho)\|^2_{\ell_2}
\end{align}
over quantum states $\rho$ and calibration vectors $\xi$ using an alternating gradient descent (AGD) or block-coordinate descent optimization.
Here, $\hat y$ is the experimentally measured data and  $\norm{\cdot}_{\ell_2}$ denotes the Euclidean norm. 
Importantly, we enforce that $\rho$ is a valid pure quantum state, and the entries of $\xi$ take only allowed values, for example, values between $0$ and $1$ for calibration parameters that are probabilities. 
These structural assumptions about the signal allow us to distinguish different sources of error and thus to solve the blind calibration problem. 
The idea of AGD minimization is to alternate gradient steps with respect to $f(\rho,\xi)$ over $\rho$, while fixing $\xi = \xi_k$, and over $\xi$ while fixing $\rho = \rho_k$ to the most recent results of a gradient step, $\xi_k$ and $\rho_{k}$, respectively.  
We perform Riemannian gradient descent with respect to the manifold of pure quantum states~\cite{roth_semi-device-dependent_2023} and Euclidean gradient-descent for the calibration vectors, respectively, and post-project onto the set of valid signal assignments after each gradient step, as detailed in the following.

\subsection{Solving blind calibration}

This section describes the alternating gradient descent (AGD) optimization algorithm for solving bilinear calibration problems over quantum states $\rho$ and calibration vectors $\xi$.
The algorithm minimizes \cref{eq:tomography least squares} via an iterative procedure based on gradient descents on Riemannian manifolds.
Each iteration alternates between a gradient step with respect to $\rho$ for fixed $\xi$ and a gradient step with respect to $\xi$ for $\rho$ constant.
In both cases, the fixed parameter is set to the most recently obtained optimal solution.

The optimization of $\rho$ is performed on the manifold of rank-$r$ positive semi-definite matrices with unit trace, while $\xi$ is constrained according to the measurement error model.
The algorithm incorporates techniques from geometric optimization, such as tangent space projections. Let $\rho_k, \xi_k$ be the values after step $k$ of the algorithm

\paragraph{$\rho$ step.}
The iteration starts by computing the Euclidean outer gradient $G_\xi = \mathcal A_{\xi_k}^\dagger(\hat y - \mathcal A(\rho_k,\xi_k))$, where $A_\xi(\rho) = A(\rho,\xi)$ denotes the linear map acting on $\rho$, of the least-squared deviation \eqref{eq:tomography least squares}.
The algorithm projects this gradient onto the tangent space of the manifold of rank-$r$ matrices.
This projection ensures that updates remain consistent with the manifold structure while accelerating convergence.
The tangent space projection at a point $\rho$ is computed as follows:
Let $\rho=U\Lambda U^\dagger$ be the eigenvalue decomposition of $\rho$, where $\Lambda$ is diagonal and the eigenvalues are arranged in decreasing order.
Let $U^{(r)}$ denote the matrix containing the first $r$ eigenvectors of $U$.
The projection of the gradient is  $P_{\mathcal T}(G) = G - (\mathbb I - P_U)G(\mathbb I - P_U)$, where $P_U=U^{(r)}(U^{(r)})^\dagger$.

The algorithm calculates the step width  
$\mu_k =\|P_{\mathcal T}(G)\|_F^2 / \|\mathcal A(\xi_k, P_{\mathcal T}(G))\|_F^2$ and 
updates the state $\rho' = \rho_k + \mu_k P_{\mathcal T}(\mathcal A_{\xi_k}^\dagger(\hat y - \mathcal A_{\xi_k}(\rho_k)))$. 
The result of the gradient step is then projected onto the set of rank-$r$ positive semi-definite (PSD) matrices.
The fixed-rank PSD projection is implemented as follows:
\begin{enumerate}
  \item Project onto the set of Hermitian matrices as  $\rho'' = (\rho' + \rho'^\dagger)/2$.
  Let $\rho''=U\Lambda U^\dagger$ be the eigenvalue decomposition of $\rho''$ with unitary $U$ and diagonal $\Lambda$.

  \item \label{step:2}
  Project onto the set of rank-$r$ PSD matrices as $\rho^{(r)}=U\Lambda^{(r)} U^\dagger$, where $\Lambda^{(r)}$ contains only the $r$ largest nonnegative entries of $\Lambda$.  
\end{enumerate}

\paragraph{Normalization of $\rho$.}
After the projection step \ref{step:2}, $\rho^{(r)}$ remains an unnormalized PSD matrix. We observe that there is a multiplicative gauge freedom between $\xi$ and $\rho$, i.e., the objective function \eqref{eq:tomography least squares} is invariant under the gauge transformation 
\begin{align}
\label{eq:gauge}
  (\xi, \rho) \mapsto (\xi/c, c\rho) 
\end{align}
for any scalar $c$. We fix this gauge by setting $\rho_{k+1} = \rho^{(r)}/\tr[\rho^{(r)}]$, and updating $\xi \mapsto \xi_{k + \frac 12} = \tr[\rho^{(r)}] \xi_k$. 
In blind calibration we additionally allow for a global sign flip when projecting onto rank-$r$ PSD matrices, since that sign can be absorbed in $\xi$. 

\begin{figure}
    \includegraphics{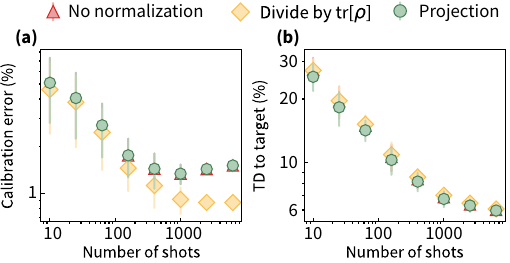}
    \caption{\textbf{Comparison of normalization methods.} Using simulated data (1000 shots per basis) from a noisy three-qubit GHZ state with 1\% Pauli noise added to each gate and miscalibration using the nine calibration parameters $\xi_{\text{actual}}$ \eqref{eq:xi actual}, we perform blind calibration using three different state-normalization techniques: No normalization (triangles), dividing by the trace of the state (diamonds), and projection (circles). The resulting calibration error is shown in (a). Using these parameters, we perform calibrated state tomography, again comparing the aforementioned normalization methods. The trace distances to the target state are shown in (b). Shifting the trace of $\rho$ into $\xi$ performs better for blind calibration, while the projection \eqref{eq:PSDTN projection} performs better for calibrated state tomography.}
    \label{fig:calibration normalization}
\end{figure}

We note that in step \ref{step:2}, one could alternatively directly project onto the set of (trace-normalized) quantum states by finding the number $\lambda \in \mb R$ that satisfies 
\begin{align}
\label{eq:PSDTN projection}
  \tr[ [\Lambda - \lambda]^{+,r}] = 1, 
\end{align}{}
where $[a]^{+,r} $ denotes the entrywise projection of a tensor $a$ onto its $r$ largest nonnegative entries \cite{smolin_efficient_2012,guta_fast_2020}.
This can be achieved, for instance, via a water-filling algorithm \cite{smolin_efficient_2012}. 
However, this projection alters the objective function and does not exploit the gauge freedom \eqref{eq:gauge}. 
Indeed, we find, that our method of `shifting' the trace of $\rho^{(r)}$ into $\xi$, performs better for blind calibration, see \cref{fig:calibration normalization}. 
However, the projection \eqref{eq:PSDTN projection} is used when we perform calibrated state tomography with a fixed calibration vector $\xi$, see Appendix~\ref{app:tomographies}. 

\paragraph{$\xi$ step.}
For the optimization of $\xi$, we have the gradient $G_{\rho} = \mc A_\rho^\dagger(\hat y - \mc A_\rho(\xi))$ with $\mc A_\rho(\xi) = \mc A(\rho, \xi)$. 
The step width $\nu_k$ for $\xi_k$ is determined using the same formula as above, omitting a tangent space projection.
The updated vector $\xi' =\xi_{k+\frac 12}+\nu_k \mathcal A^\dagger_{\rho_{k+1}}(\hat y - \mathcal A_{\rho_{k+1}}(\xi_{k+\frac 12}))$ is then projected onto the set of valid calibration parameters via the projection $\mc P_C$ yielding $\xi_{k+1} = \mc P_C (\xi')$ .

\paragraph{Termination criteria}
The algorithm continues until the relative deviation of the objective function satisfies
$\|\hat y - \mathcal A(\xi,\rho)\|_2/\|\hat y\|_2\leq\varepsilon$, where $\varepsilon$ is a predefined threshold, or until a maximum of iterations is reached. Here we have used $\varepsilon=10^{-2}$ and 100 maximum iterations.

\paragraph{Initialization}
When performing blind calibration, we initialize $\rho$ and $\xi$ by randomly sampling from a specified region around the anticipated values. For $\xi$, we have chosen a region within 15\% of the directly measured values, while for $\rho$, we have chosen a region within 10\% of the target state. 

\begin{figure}
    \centering
    \includegraphics{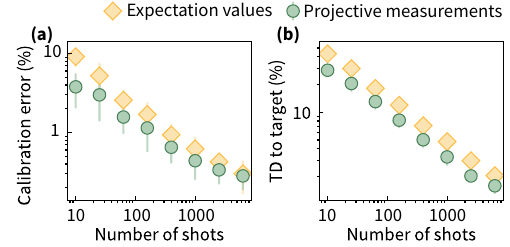}
    \caption{\textbf{Comparison of projective measurement and expectation value calibrations}. Using simulated data (1000 shots per basis) from a noiseless three-qubit GHZ state and the nine calibration parameters $\xi_{\text{actual}}$ \eqref{eq:xi actual}, we perform blind calibration using projective measurements (circles) and expectation value measurements (diamonds). The calibration error is shown in~(a). The recovered parameters are then used to perform calibrated state tomography, for which the TD to the target state is shown in~(b). 
    }
    \label{fig:projective vs expectation}
\end{figure}

\paragraph{Expectation values versus projective measurements}
In blind calibration, we use projective measurement data, compared to expectation values as elaborated in Ref.~\cite{roth_semi-device-dependent_2023}. 
In the finite-sample regime, this more fine-grained data yields better recoveries of both the calibration parameters and the quantum state preparations, as demonstrated in \cref{fig:projective vs expectation}. 
We observe that both for the calibration error and the TD to the target state, the projective measurements outperform the expectation value measurements, until a saturation point is reached where they reach equal performance.

\section{Calibrated state tomography}
\label{app:tomographies}

In order to assess the quality and compare different calibrations, we use those calibrations to obtain tomographic state estimates from the same type of measurement data (projective Pauli measurements) on different states. 
To do this, we perform what we call `calibrated tomography'. In calibrated tomography, we fix a calibration $\xi$ in our measurement model \eqref{eq:linear measurement model appendix} and run Riemannian gradient descent on the set of quantum states in order to obtain the least-square fit (the minimizer of \cref{eq:tomography least squares}) for $\rho$.
Specifically, we run the same algorithm as in blind calibration, outlined in Appendix~\ref{app:algorithm}, but rather than alternating the gradient steps, we fix the calibration $\xi$. 
At the end of the algorithm, we project onto the set of quantum states using the projection \eqref{eq:PSDTN projection} \cite{guta_fast_2020}. 

Importantly, while in blind calibration we want to exploit all available structural assumption in order to obtain the best possible calibration and therefore restrict to rank-$1$, when assessing a calibration, we want to obtain the most accurate state estimate, and thus remove the rank constraint, setting $r =D$ to be the Hilbert-space dimension $D$.

In our comparisons, we perform calibrated tomography using the ideal calibration ($\xi = (1, 0, \ldots)$), corresponding to \emph{standard tomography}, the directly measured estimate, and the blind calibration estimate. 

\begin{figure}
    \centering
    \includegraphics{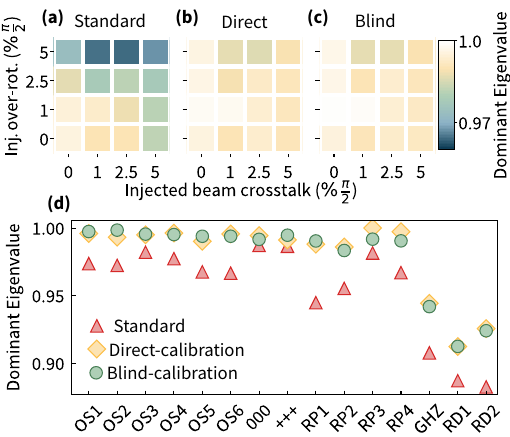}
    \caption{\textbf{Dominant eigenvalues of calibrated state recoveries from experimental data. }
    States are recovered using full-rank standard, direct-calibration, and blind-calibration tomography for data with injected errors (a) and native errors (b). The dominant eigenvalues are close to $1$, showing that the the states are near-pure, and justifying the  rank-1 constraint in blind-calibration.}
    \label{fig:dominant-eigenvalues}
\end{figure}

In \cref{fig:dominant-eigenvalues}, we show that the rank-$1$ constraint imposed in blind calibration is  approximately true for the experimental data by comparing the dominant eigenvalue of the state estimate using calibrated tomography with different calibration parameter estimates. 
It becomes apparent that miscalibrations lead to a higher weight on subdominant principal components of the quantum state while the direct and blind calibrations have significanly larger dominant eigenvalues that are---as expected---close to 1 for product states, and somewhat smaller for states prepared by circuits involving $2$-qubit gates. 

\begin{figure}
    \includegraphics{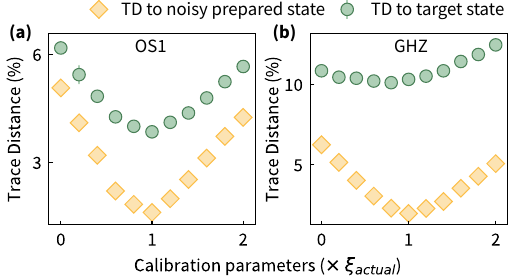}
    \caption{\label{fig:trace-distance-verification}
    \textbf{Evidence for validity of trace distance to the target state.}
    Comparison of the state estimate obtained from calibrated tomography with calibration parameters $(1-c) \xi_{\text{ideal}} + c \xi_{\text{actual}}$ for $c \in [0,2]$ with $\xi_{\text{ideal}} = (1,0, \ldots)$ and $\xi_{\text{actual}}$ given in \cref{eq:xi actual} using $1000$ simulated shots per measurement basis from from noisy state preparations using a Pauli noise channel with strength 1\% per gate for the probe state OS1 (a) and the GHZ state (b).  
    Shown are the TD to the noisy (simulated) state preparation (diamonds) and the target state (circles). Error bars are one standard deviation.
    }
\end{figure}

The quality of different calibration can be measured in the trace distance of the calibrated-tomography estimate to the true state preparation. 
But while in simulations we can compute this quantity since the true (simulated) state preparation is known, this information is not available in experiments. 
This raises the question what a good measure of quality for the different calibrations is. 
Here, we use the trace distance (TD) \emph{to the target state}, that is, the intended state preparation as a substitute for the trace distance to the true experimental state preparation.
The intuition behind this choice is that a lower TD to the target state indicates a better calibration, since errors in the calibration estimate would have to conspire with the noise in the state preparation in order to yield a lower value compared to a more accurate calibration estimate. 
This expectation is confirmed in simulations of noisy state preparations. 
In \cref{fig:trace-distance-verification}, we show both the TD to the target state and to the noisy preparared state from which the data was generated as we scale the calibration from the ideal calibration $\xi_0 = (1, 0, \ldots)$ to the actual calibration of the data set $\xi_{\text{actual}}$ and beyond. 
As expected, the TD to the target state attains a minimum value at the actual calibration parameters, and generally correlates with the trace distance to the true state preparation. 
We also find that the amount of correlation differs for different states. 
For example, for the GHZ state there seems to be a weaker correlation than for one of our probe states. 

\section{Error models}
\label{app:error-models}

\label{sec:overrotation}

\subsection{Overrotations}

The transformed Pauli bases resulting from an overrotation can be derived through transforming the $Z$ basis by an over-rotated gate unitary. For the $X$ basis, we rotate by $R_y(-\frac{\pi}{2} - \xi_\text{\rm OR} \frac{\pi}{2})$, where $\xi_\text{\rm OR} \frac{\pi}{2}$ is assumed to be a small overrotation angle. This yields

\begin{align}
    X \mapsto& R_y\left(\frac{\pi}{2} + \xi_\text{\rm OR} \frac{\pi}{2}\right) Z R_y\left(-\frac{\pi}{2} - \xi_\text{\rm OR} \frac{\pi}{2}\right)\\
= & \cos(\xi_\text{\rm OR}\pi)X - \sin(\xi_\text{\rm OR}\pi)Z.
\end{align}
Assuming the extra angle $\xi_\text{\rm OR} \frac{\pi}{2}$ is small, we therefore obtain
\begin{align}
    X &\mapsto X - \xi_\text{\rm OR} Z
\end{align}

\noindent The same derivation applies for the $Y$ basis.

\label{sec:beam-crosstalk}
\subsection{Beam crosstalk}

During single-qubit rotations on a target ion, light from the laser beam  may cause unwanted illumination of neighboring ions as shown in \cref{fig:intro-fig}(e) causing a rotation by some angle $\xi_{l} \cdot \pi/2$ on the left ion and $\xi_{r} \cdot \pi/2$ on the right ion.
Due to the laser's potentially non-uniform phase front in at the plane of the ions, such crosstalk may also introduce unintended phase differences $\phi_l$ and $\phi_r$ between the target qubit and its left and right neighbors.

Unlike overrotations, beam crosstalk transformations do not have a simple form. Instead, we must determine the unitary acting on the qubits as a result of both the gates on the target qubits as well as the induced rotations on the neighboring qubits.  The resulting basis transformation is determined by applying this unitary to the  $Z$ basis. For two qubits, the gates consist of the target qubit rotation and the crosstalk ``gate" on the neighboring qubit(s). For example, ignoring any phase difference between target and neighbor, measuring in the $XY$ basis results in the crosstalk circuit

\begin{center}
    \begin{quantikz}
    &\gate{R_Y(-\frac{\pi}{2})}&\gate[style={fill=red!10}]{R_X(\xi_l \frac{\pi}{2})}& \meter{} \\
    &\gate[style={fill=red!10}]{R_Y(-\xi_r\frac{\pi}{2})}&\gate{R_X(\frac{\pi}{2})}& \meter{} &
    \end{quantikz}
\end{center}
for which we can construct the unitary, 

\begin{align}
    U_{bc} &= R_y^{(1)}\left(-\frac{\pi}{2}\right) R_y^{(2)}\left(-\xi_r\frac{\pi}{2}\right) R_x^{(1)}\left(\xi_l\frac{\pi}{2}\right) R_x^{(2)}\left(\frac{\pi}{2}\right) .
\end{align}
We can apply this unitary transformation to the $Z$ basis as

\begin{align}
    Z &\mapsto U_{bc} Z U^{\dagger}_{bc}
\end{align}
and extract the Pauli terms which are first order in $\xi$, resulting in the transformation, 

\begin{align}
    XY &\mapsto XY + \xi_l \frac{\pi}{2} YY.
\end{align}

Now, including the phase difference between the neighbor and target requires the use of arbitrary rotations, rather than strictly $R_X$ and $R_Y$ gates on the neighbors. The arbitrary rotations applied to the left and right neighbors have a phase equal to that of the target phase plus $\phi_l$ or $\phi_r$. 

\begin{multline}
    XY \mapsto XY + \xi_l \pi \cos(\phi_l)YY + \\ \xi_l \pi \sin(\phi_l)ZY - \xi_r \pi  \sin(\phi_r) XZ.
\end{multline}

For a system with a purely Gaussian beam and perfectly matched beam spacing to ion spacing, the beam crosstalk will be symmetric to the left and right, reducing the necessary parameters from two angles and two phases down to a single angle and single phase. Additionally, we may assume a uniform phase front across the laser beam, resulting in only a single error parameter corresponding to the crosstalk magnitude. Conversely, for a system with large beams or far-reaching aberrations, we can expand our model to include next-nearest neighbor crosstalk, adding an additional angle and phase for each neighbor we consider.

\label{sec:classical-readout}

\subsection{Bright and dark errors}

Unlike rotation errors, it is intuitive to model readout errors as a map $\Lambda$, acting on the projective measurements.
For dark and bright errors, we can directly write the complete map:

\begin{align}
    \proj{0}_M &\mapsto (1-p_0) \proj{0} + p_0\proj{1} \\
    \proj{1}_M &\mapsto p_1 \proj{0} + (1-p_1)\proj{1} 
\end{align}

\subsection{Detector spillover}
\label{sec:pmt-spillover}

During the fluorescence detection process, photons emitted by individual ions are collected and counted on individual channels of a detector array. After a fixed detection time, a threshold for the photon number is used to discriminate ions in state $\ket{0}$ and state $\ket{1}$. Due to imperfect alignment of the detector and electrical crosstalk between the channels, photon counts from a bright ion may spill over to neighboring channels as shown in \cref{fig:intro-fig}(d), potentially resulting in a dark ion being classified as bright. 

Detector spillover errors occur with probabilities $p_{\text{left}}$ to the left and $p_{\text{right}}$ to the right. Unlike the previous errors, detector spillover is a purely multi-qubit error, and therefore the simplest map $\Lambda$ describing detector spillover is the two-ion case:

\begin{align}
    \proj{00}_M &\mapsto \proj{00}_M\\
    \proj{01}_M &\mapsto  (1-p_{\text{left}})\proj{01}_M + p_{\text{left}} \proj{11}_M \\
    \proj{10}_M &\mapsto  (1-p_{\text{right}})\proj{10}_M + p_{\text{right}} \proj{11}_M\\
    \proj{11}_M &\mapsto  \proj{11}_M
\end{align}

where we can see that spillover only occurs when a bright ion has a dark ion as a neighbor.

\subsection{Combining errors}

To construct the final measurement map $\mc A$, the transformations due to individual errors on individual or pairs of qubits must be combined. 
This will lead to higher-order effects, which we neglect, keeping only terms that are linear in the calibration parameters.

\subsection{Parameter values for numerical benchmarks}

In our numerical benchmarks we fix the values of the calibration parameters to $\xi_{\text{actual}}$ with components
\begin{align}\label{eq:xi actual}
\xi_\text{\rm OR} &=1 \% \\
\notag p_0 &=0.32 \%\\
\notag p_1 &=1.541 \%\\
\notag p_\text{left} &=0.17 \%\\
\notag p_\text{right} &=0.41 \%\\
\notag \xi_l &=2.56 \%\\
\notag \xi_r &=1.18 \% \\
\notag \phi_l &=\frac{\pi}{4}\\
\notag \phi_r &=\frac{\pi}{8}
\end{align}

\section{Native system calibration}
\label{app:native calibration}

\subsection{Blind calibration}

\label{app:optimal states}
\begin{figure}
    \centering
    \includegraphics{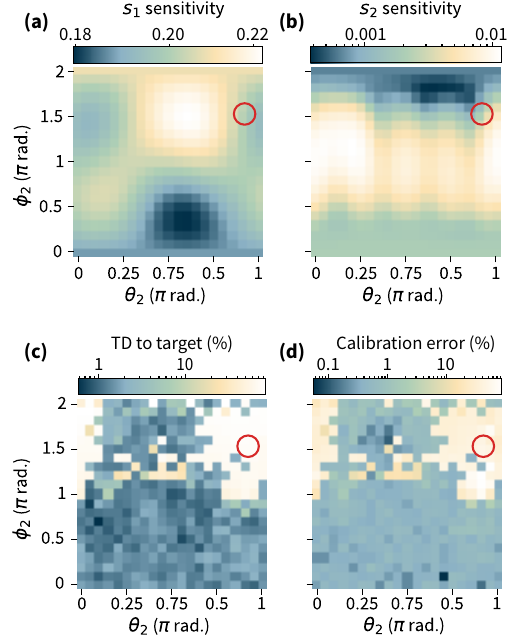}
    \caption{\textbf{State sensitivity tests} for a noiseless three-qubit random product state, RP1, simulated with miscalibrations $\xi_{\text{actual}}$ \eqref{eq:xi actual}. The two outer qubits are held at fixed angles, while the middle qubit is scanned through all possible states. The trace distance to the target and the calibration error presents complex features. The product state is shown by the red circle, and is seen to be particularly poor.
    }
    \label{fig:sensitivity}
\end{figure}

\subsubsection{Optimal states}

We observe that some states do not provide accurate calibrations for particular error types. We predict that this behavior results from the state's lack of sensitivity to the error. We predict two cases in which the probe state may give bad results.

\begin{itemize}
    \item[1.] The measurement data for the state is insensitive to certain measurement errors. In this case, the ideal measurement, $\mc A_0(\rho)$ is approximately equal to the calibrated model, $\mc A(\xi, \rho)$.

    \item[2.] The measurement data are sensitive to the measurement error, but the optimal solution to the uncalibrated recovery is still a valid quantum state, that is, $\mc A^{-1}(y) \in \mc D(r)$. In this case, the density matrix constraint is not sufficient to distinguish a measurement error from a state-preparation error, resulting in a plateau in the optimization landscape.
\end{itemize}

We propose two methods for calculating these sensitivities:

\begin{itemize}
    \item[1. ] $s_1(\xi, \rho) \coloneqq \norm{\mc A(\xi, \rho) - \mc A_0(\rho)}$
    \item[2. ] $s_2(\xi, \rho) \coloneqq  \norm{\mc A_0^{-1}(\mc A(\xi, \rho)) - \mc P_{\mc D(r)} [\mc A_0^{-1}(\mc A(\xi, \rho))]}$
\end{itemize}

We test whether these proposed sensitivity measurements can predict the performance of the blind calibration by simulating three-qubit product state recoveries and sensitivities. We vary the angle of the center qubit over the Bloch sphere, measuring $s_1$ and $s_2$ and performing a blind calibration for each unique state. We find that the proposed sensitivity measurements $s_1$ and $s_2$ are not correlated to the performance of the blind calibration as shown in \cref{fig:sensitivity}. 

Therefore, in order to find an optimal set of states, we look only to the performance of the blind calibration. We optimize over three-qubit product states defined by ($\theta_i$, $\phi_i$), for each $i$ qubit. For each state, we compute an objective which takes into account the calibration error at each value over the range of expected $\xi$. First, we minimize the average calibration error over the range. Second, we minimize the maximum total calibration error. We perform this optimization with and without measurement shot noise, and with a normalized and non-normalized calibration error measurement. This results in six optimal probe states defined by the azimuthal and polar angles as ($\theta_1$, $\phi_1$, $\theta_2$, $\phi_2$, $\theta_3$, $\phi_3$) in units $\pi$ radians:

\begin{itemize}
    \item[1. ] \textbf{OS1} with angles (0.910, 1.978, 0.475, 0.378, 0.467, 0.480)
    \item[2. ] \textbf{OS2} with angles (0.463, 0.846, 0.742, 1.618, 0.530, 0.726)
    \item[3. ] \textbf{OS3} with angles (0.0316, 0.259, 0.374, 1.820, 0.372, 0.795)
    \item[4. ] \textbf{OS4} with angles (0.993, 1.954, 0.494, 0.999, 0.0767, 1.521)
    \item[5. ] \textbf{OS5} with angles (0.870, 1.967, 0.406, 1.012, 0.426, 0.548)
    \item[6. ] \textbf{OS6} with angles (0.352, 1.380, 0.707, 0.243, 0.448, 0.219)
\end{itemize}
\subsubsection{Test states}

We use nine states to test the blind calibration. For product states, we define the azimuthal and polar angles for each qubit as: ($\theta_1$, $\phi_1$, $\theta_2$, $\phi_2$, $\theta_3$, $\phi_3$) in units $\pi$ radians. The test product states are

\begin{itemize}
    \item[1. ] \textbf{$\ket{000}$} with angles (0, 0, 0, 0, 0, 0)
    \item[2. ] \textbf{$\ket{+++}$} with angles (0.5, 0, 0.5, 0, 0.5, 0)
    \item[3. ] \textbf{RP1} with angles (0.871, 1.427, 0.713, 1.190, 0.693, 1.477)
    \item[4. ] \textbf{RP2} with angles (0.723, 1.198, 0.924, 1.533, 0.871, 0.485)
    \item[5. ] \textbf{RP3} with angles (0.736, 0.559, 0.654, 0.422, 0.783, 1.211)
    \item[6. ] \textbf{RP4} with angles (0.957, 0.105, 0.942, 0.270, 0.704, 0.773)
\end{itemize}

We also test three entangled states. Entanglement is generated using a variant \cite{xxgates} of the Molmer-Sorensen (MS) gates \cite{sorensen_quantum_1999}, parameterized by the angle $\chi_{ij}$ on qubits $i$ and $j$, where $\chi_{ij}=\frac{\pi}{4}$ produces a fully-entangled state.

\begin{itemize}
    \item[1. ] \textbf{GHZ}, the Greenberger-Horne-Zeilinger state $\ket {000} + \ket {111}$.
    \item[2. ] \textbf{RD1} with single-qubit rotations by angles (0.872, 0.426, 0.714, 0.190, 0.693, 0.477), then fully entangling MS gates on each qubit pair.
    \item[3. ] \textbf{RD2} with single-qubit rotations by angles (0.722, 0.198, 0.923, 1.533, 0.871, 0.485), then fully entangling MS gates on each qubit pair.
\end{itemize}

\begin{figure}
    \centering
    \includegraphics{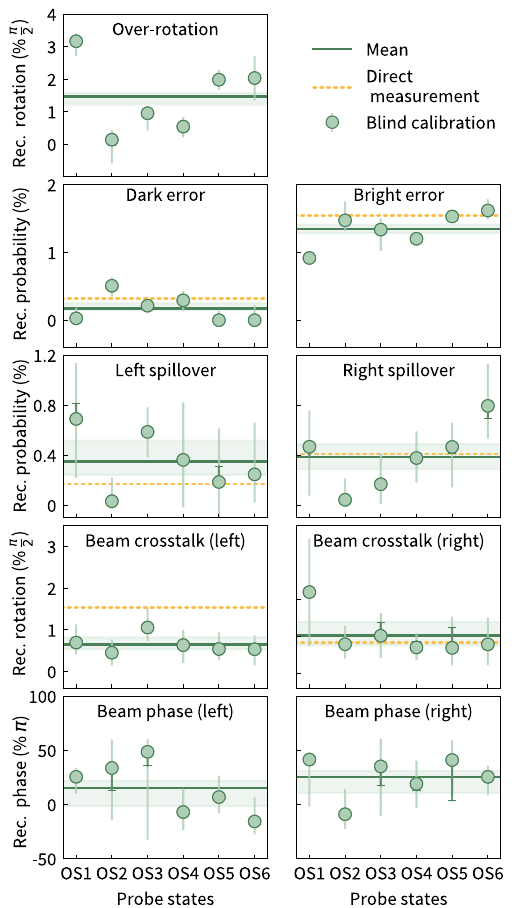}
    \caption{\textbf{Complete set of native calibration parameters} of the experimental system recovered using blind calibration and direct calibration. The recovered values for each state (circles) and their averages (solid line) are compared to the direct experimental measurement (dashed line) where available.}
    \label{fig:all-recoveries}
\end{figure}

\begin{figure}
    \centering
    \includegraphics{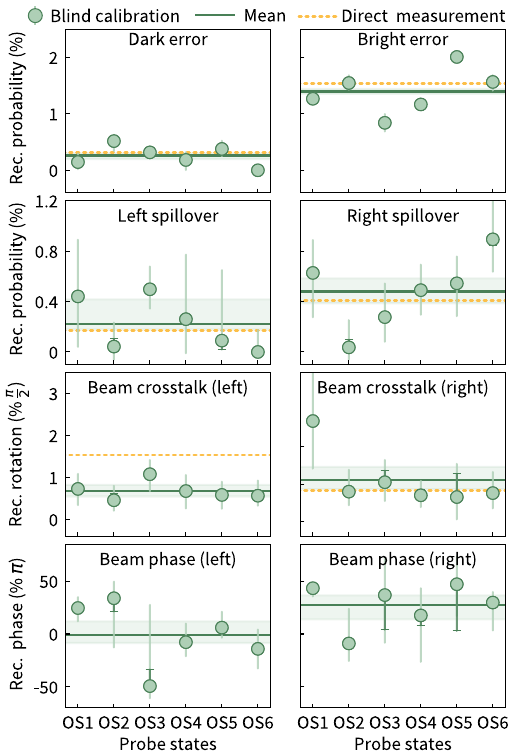}
    \caption{\textbf{Native calibration parameters excluding overrotation errors} recovered using blind calibration and direct calibration. The recovered values for each state (circles) and their averages (solid line) are compared to the direct experimental measurement (dashed line) where available.}
    \label{fig:recoveries-no-overrotation}
\end{figure}

\subsection{Full recovery}
\label{app:full recovery}

\subsubsection{Parameter choice}
\label{app:parameter-optimization}
Due to the interplay between parameter number, model accuracy, and recovery performance, we expect an optimal set of modeled calibration parameters to exist, where removing and adding parameters from the model both result in worse recovery performance. To find this optimal set for the experiment in its native setting, we perform blind calibration with various subsets of calibration parameters, taking advantage of the post-hoc nature of the method. We rely on two metrics to assess the quality of the calibration: the difference in trace distance between blind and direct tomography and the calibration error of the recovery. We find that removing overrotation errors from the model results in an optimal recovery in both trace distance and calibration error, shown in \cref{fig:parameter-removal}. We therefore include only eight calibration parameters in the experimental calibration: classical readout errors, detector spillover errors, beam crosstalk magnitude, and beam crosstalk phase.

\subsubsection{Recovery}

We compare all measurements (direct and blind) of the native calibration parameters for a model including overrotation (\cref{fig:all-recoveries}) and a model excluding overrotations (\cref{fig:recoveries-no-overrotation}).
We find good agreement between the parameters obtained in blind calibration and the directly measured values where available. An exception is the left beam crosstalk parameter $\xi_{l}$, where the directly measured value is inconsistent with the blind  calibration estimates, which are consistent across the different probe states. 

\begin{figure}
    \centering
    \includegraphics{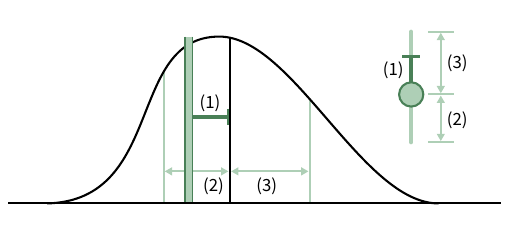}
    \caption{\textbf{Error bars for biased data.} A statistically sampled distribution is shown (black curve) with the reported value (wide bar). The reported value is not equal to the median of the distribution, so we report two error bars. The distance from the reported value to the median of the distribution is represented with a single capped error bar (1). One standard deviation of the sampled distribution is taken to approximate the lower and upper error bars (2, 3) for the reported value.}
    \label{fig:error-bar-graphic}
\end{figure}

\subsubsection{Error bars}
\label{app:error-bars}

We perform statistical sampling to obtain error bars for our reported values. By sampling from the distribution of experimental data and performing blind calibration on each sample, we get an estimate for the output distribution. We find that our recovery is biased for some parameters, resulting in an output distribution that is not centered on our reported value. Although the mean of the output distribution may not be equal to our value, we assume the width of the distribution is a good approximation for the true distribution. We use a light error bar without caps to show this width (one standard deviation) obtained from the sampled distribution. We use a second darker, capped error bar to show the distance from our recovered value to the median of the sampled distribution. The distribution and corresponding plotted point are shown pictorially in \cref{fig:error-bar-graphic}.

\subsection{Direct measurements}
\label{app:direct measurement}

\subsubsection{SPAM}
\label{app:spam-measurement}

We independently measure the state preparation and measurement (SPAM) errors in our system. We construct a SPAM matrix, which we can invert and apply to the measured populations to infer the SPAM-free results. Through this method, we also directly measure the dark and bright errors and the detector spillover in the system.

To construct this SPAM matrix, we use a single ion and adjust the trap voltages to move the ion into the locations the three ions occupy when trapped together. We prepare the ion either in the dark or in the bright state, the latter with a high-fidelity microwave pulse, before performing a state measurement.  For each ion position, we measure the probability of a bright ion on each of the three detector channels. The probability of measuring the ion as bright at its position gives us the bright error, while the probability of measuring a bright ion at the unoccupied positions gives us the detector spillover. Similarly, we prepare the ion in its dark state to characterize the dark error.

From this information, we can reconstruct the probability of measuring any state for any prepared state without Raman addressing error, giving us a complete readout matrix, which we use to correct the data.

\subsubsection{Sensitivity to noise}
\label{app:direct-sensitivity}

We can calculate the dependence of readout error estimates on noise by directly calculating the mapping from prepared to measured states. For simplicity, we will derive the dependence for a single-qubit readout error, with dark error probability $p_0$ and bright error probability $p_1$. The ideal channel without depolarizing noise will be:

\begin{align}
    \proj{0}_M &\mapsto (1-p_0)\proj{0} + p_0 \proj{1}\\
    \proj{1}_M &\mapsto (1-p_1)\proj{1} + p_1 \proj{0}
\end{align}

When measuring this ion on its detector channel, we have a $1-p_0$ ($1-p_1$) chance of measuring it as dark (bright) and a $p_0$ ($p_1$) chance of measuring it as bright (dark). We perform many experimental shots to get as estimate for $p_0$ ($p_1$). However, when depolarizing noise with strength $\lambda$ is present:

\begin{align}
    \proj{0}_{\text{depol.}} &= (1-\frac{\lambda}{2})\proj{0} + \frac{\lambda}{2} \proj{1}\\
    \proj{1}_{\text{depol.}} &= (1-\frac{\lambda}{2})\proj{1} + \frac{\lambda}{2} \proj{0}
\end{align}

resulting in a readout characterized by:

\begin{align}
    \proj{0}_{M} &= ((1-\frac{\lambda}{2})(1-p_0) + \frac{\lambda}{2} p_1)\proj{0} \\
    &+ (\frac{\lambda}{2} (1-p_1) + (1-\frac{\lambda}{2})p_0) \proj{1}\\
    \proj{1}_{M} &= ((1-\frac{\lambda}{2})(1-p_1) + \frac{\lambda}{2} p_0)\proj{0} \\
    &+ (\frac{\lambda}{2} (1-p_0) + (1-\frac{\lambda}{2})p_1) \proj{1}\\\end{align}

When measuring dark and bright statistics on the target detector channel, we will estimate $p_0$ and $p_1$ as

\begin{align}
    \tilde{p_0} &= p_0 + (\frac{1-p_0-p_1}{2}) \lambda\\
    \tilde{p_1} &= p_1 + (\frac{1-p_0-p_1}{2}) \lambda
\end{align}

from which it is clear that the difference between the actual values $p_0$ \& $p_1$ and our estimated values $\tilde{p_0}$ \& $\tilde{p_1}$ scales linearly with depolarizing strength, $\lambda$. A similar calculation can be performed for spillover errors.

\subsubsection{Beam crosstalk}
\label{sec:beam-crosstalk-measurement}

To measure the beam crosstalk present in the system, we prepare a chain of three ions in the state $\ket{000}$. We apply a single individual target beam and measure the resulting Rabi flop on each qubit, including the target. We calculate the ratio between the target Rabi frequency and the neighboring Rabi frequencies for each ion pair, giving us the magnitude of beam crosstalk. We find that the nearest-neighbor crosstalk magnitudes are asymmetric in direction, but consistent across ion positions.

\end{document}